\documentclass[a4paper,alpha-refs]{aejstyles}

\journal{aej}

\usepackage{graphicx}
\usepackage{siunitx}
\usepackage{caption}
\usepackage{subcaption}
\usepackage{hyperref}
\usepackage{url}
\usepackage{amsmath}
\usepackage{multirow}
\usepackage[flushmargin]{footmisc}

\title{Rising Stargirls: Benefits of a Creative Arts-Based Approach to Astronomy Education for Middle-School Girls from Underrepresented Groups}

\author[1,\authfn{1}]{Maya Silverman}
\author[1,\authfn{1}]{Aomawa L. Shields}
\author[1,2]{Jessica N. Howard}
\author[1]{Vidya Venkatesan}
\author[1,3]{Kiana Whitfield}

\affil[1]{Department of Physics and Astronomy, University of California, Irvine, CA 92697 USA}
\affil[2]{Kavli Institute for Theoretical Physics, University of California, Santa Barbara, CA 93106 USA}
\affil[3]{University of Maryland, College Park, MD 20742 USA}

\authnote{\authfn{1}msilver2@uci.edu; shields@uci.edu}

\papercat{Research Article}

\runningauthor{Silverman et al.}

\jvolume{04}
\jnumber{1}
\jyear{2024}

\begin{document}

\begin{frontmatter}
\maketitle
\begin{abstract}

Women from historically marginalized groups in the sciences continue to be severely underrepresented in the fields of physics and astronomy. Young girls identifying with these groups often lose interest in science, technology, engineering, and math (STEM) fields well before college. Middle school (grades 6-8) emerges as a pivotal phase for nurturing science identities among girls. The educational program Rising Stargirls offers creative arts-based astronomy workshops for middle-school girls, with the aim of cultivating their science identities. We retrospectively analyze participants' responses to four key assessment items through which their engagement in science and their science identities before and after the workshops are assessed. Our findings overwhelmingly indicate that girls exhibit heightened engagement in science and enhanced science identities after engaging in the Rising Stargirls program. These outcomes underscore the merits of fostering creativity and integrating the arts into science education.

\end{abstract}

\begin{keywords}
Science identity; Astronomy education; Creative arts; Underrepresented Minority Students (URM) 
\end{keywords}

\end{frontmatter}

\section{Introduction}

Women from historically marginalized groups in the sciences (American Indian or Alaska Native, Black or African-American, Hispanic or Latinx, Native Hawaiian or other Pacific Islander, hereafter ``URM'') continue to be severely underrepresented in the fields of physics and astronomy. This remains true across all levels of the academic hierarchy in physics and astronomy, and disparities persist in advances across URM groups. The number of Latinx women earning bachelor’s degrees in astronomy has doubled to 16\% since 2002, and though still under the Hispanic/Latinx population of 19.1\% in the United States, progress has been made over the past 20 years \citep{porter:2019}. In contrast, the number of African American women earning bachelor’s degrees in astronomy has remained relatively static over the same time period, at 3\%, despite African Americans comprising 13.6\% of the United States population \citep{Mulvey:2020}. At the PhD level, during the 2018-19 academic year, of the 1,875 total doctorates earned in physics, ten were awarded to Latinx women, and one was awarded to an African American woman \citep{AIPfig}. 

Girls from URM groups often stop pursuing their interest in science, technology, engineering, and math (STEM) fields long before they enter college due to a lack of self-confidence and few role models who look like them \citep{Weir2007}. The combined impact of internal and external forces on the educational trajectories of women of color underscores that there is a small window of opportunity available to cultivate interest and nurture the involvement of young URM girls in astronomy and physics. The \textit{Planetary Science and Astrobiology Decadal Survey 2023-2032} has identified that ``increasing diversity and representation in NASA Planetary Science Division missions requires a concerted effort to engage members of underrepresented communities at early stages of career/education'' \citep{national_academies_of_sciences_origins_2022}.

Middle school is often the time where girls (aged 10-14) start to place a significant amount of focus on their physical appearance, and shift the focus away from their mental aptitude and accomplishments \citep{Gurian2012}. As a result, feelings of low self-esteem and a lack of self-confidence begin to take root \citep{Rakow2009, Gurian2012}. This age range is both a challenging time for young girls, and an ideal time to develop and nurture their science identities, which is deemed critical for the retention of African American students pursuing an undergraduate degree in Physics and Astronomy \citep{TeamUp}.

Studies on teaching methods and assessment strategies suggest that students’ conceptual understanding of astronomy improves as a result of the implementation of learner-centered, interactive exercises in class \citep{Alexander2005, Rudolph2010}. Additionally, the incorporation of literary exercises and role-playing exercises has been shown to improve students’ engagement and excitement in astronomy and astrobiology, as well as their confidence in answering questions \citep{Garland2007, Crider2012}. \cite{Wulf:2013} show that shifting their curriculum such that students take on an active role in their learning promotes development of student agency and communication. And according to the \citet{NRC2009}, informal settings provide an increased motivation to learn about the natural world, and to develop and nurture an identity that includes practicing and even contributing to science. In summary, learner-centered, interactive, literary, and role-playing exercises and giving students an active role in their learning have been shown to improve students' engagement with science and science identity---their perception of themselves as a ``science person''.

To add to these results, we study the effects of creative arts-based astronomy curricula on the broader science engagement and science identity of middle-school girls. To not only involve but enthrall girls in the possibility of identifying with and potentially becoming astronomy and planetary science researchers means engaging girls in the discovery and exploration of astronomical phenomena and questions, using who they are and what they think and feel about the universe as the lens through which this information is processed.

To study the impact of creative arts-based education, we analyze data from anonymous program-improvement surveys conducted by Rising Stargirls between 2015 and 2023. Rising Stargirls\footnote{\href{http://www.risingstargirls.org/}{http://www.risingstargirls.org/}}, a  program founded by Professor Aomawa Shields and facilitated by the authors, offers creative arts-based astronomy workshops for middle-school girls with the mission to build personal connections between the girls and the universe, of which they are an integral part. By integrating creative arts-based strategies such as free writing, visual art, and theater exercises, Rising Stargirls has developed a curriculum that ``addresses each girl as a whole by providing an avenue for individual self-expression and personal exploration that is interwoven with scientific engagement and discovery.''\footnote{\href{http://www.risingstargirls.org/}{http://www.risingstargirls.org/}} Using this approach, we highlight how incorporating creative arts at the core of astronomy education can enhance participants' engagement with science and strengthen their science identity.

Data analyzed in this study are in response to four survey items. Science identity is probed by analyzing participant responses to questions asking whether they believe they can do well in science and if they see themselves as scientists. Engagement in science is probed by analyzing participant responses to questions asking if they talk to their families and friends about science and if they enjoy their science classes.

This study explores the impact of creative arts-based astronomy activities on middle-school girls of all colors and backgrounds. However, given that a majority of participants identify with at least one URM group (see Table~\ref{tab:demographics}), these results also highlight the benefits of creative arts-based astronomy activities on girls from URM groups.

Although Rising Stargirls is an astronomy-focused education program, we offer two key reasons for its use in studying girls' engagement and identity in science more broadly. First, middle-school girls often lack access to formal astronomy classes in school, so gauging changes in their engagement and identity with science is more meaningful at this stage of their education. Second, our hypothesis is not confined to astronomy; it applies across the sciences. Therefore, the assessment questions evaluate participants' engagement with science and their science identity. This approach allows us to explore the benefits of creative arts-based curricula in science education, with astronomy workshops serving as our model.

In short, this work addresses the broader question: How does incorporating creative arts into science education increase engagement with science and shape science identity among middle-school girls? Using data collected by an educational outreach program that engages middle-school girls in creative arts-based astronomy workshops, we analyze responses to four survey items: two designed to assess participants' engagement in science, and two aimed at evaluating their science identity. Our results demonstrate that encouraging middle-school girls to bring their creativity to science promotes their engagement with science and nurtures their science identity.

This paper is organized as follows: Section~\ref{sec:programdescr} describes the Rising Stargirls program, including the resources that the program has made publicly available. Section~\ref{sec:methodology} details the data collection and analysis methodology. Section~\ref{sec:results} describes the results of our analysis, separated by Rising Stargirls' in-person and virtual workshops. Finally, Section~\ref{sec:discussion} discusses the impacts, limitations, and conclusions of our results.

\section{Program Description}
\label{sec:programdescr}

Rising Stargirls was founded in 2014 with the mission to build personal connections between middle-school URM girls and the universe with the aim of encouraging increased participation in astronomy and astrobiology from groups traditionally underrepresented in the sciences. It integrates scientific discovery with arts-based activities to invite girls to bring their whole selves to the science learning process. 

The curriculum is implemented in after-school or summer camp workshops for middle-school students. Rising Stargirls' workshops have been conducted in-person on the East and West Coasts of the United States, and virtually on Zoom for students across the world. In the six workshops analyzed in this study, Rising Stargirls reached 126 students across the world! The format for the workshops is flexible but the curriculum remains largely consistent. The curriculum is a collection of arts-based astronomy activities that are selected and curated from free, publicly-available Rising Stargirls educational materials (see Section~\ref{sec:resources}). 

An example arts-based astronomy activity that participants in Rising Stargirls' workshops experienced focuses on teaching girls about extrasolar planets (or "exoplanets") -- planets that orbit stars other than our Sun. The activity begins with a lecture that includes images and videos of exoplanets to give them a visual representation. During the lecture, participants have the opportunity to share what they know with facilitators and peers and ask questions. They learn the definition of a planet, and the difference between planets within our solar system and exoplanets. After being introduced to basic concepts, they apply their knowledge through a hands-on activity called, "Build Your Own Exoplanet." 

The goal of this activity is to give participants an opportunity to think critically and creatively about the types of exoplanets that may exist. They are encouraged to consider what they might look like and what elements of a planet might be conducive  or detrimental to life. Each participant receives a large sheet of construction paper, a pencil, a pen, crayons, and markers to design their own exoplanet. They get to decide whether their planet is rocky like Earth or giant like Jupiter; whether it is cold or hot. They are also asked to think about whether their planet contains land or ocean; if it's an aquaplanet, or a land planet. Additionally, they are asked to reflect on what life might look like on that planet. This activity is intended to encourage curiosity, creativity, and imagination, and help girls develop their scientific thinking.  

This example activity is supported by multiple other activities that participants do earlier in the workshop. For example, an activity that precedes this one is called "Art and the Cosmic Connection," adapted with permission from \textit{NASA’s Discovery \& New Frontiers Programs}\footnote{\href{https://eurekus.org/nasa-art-cosmic-connection}{https://eurekus.org/nasa-art-cosmic-connection}}. In this activity, girls create their own artist depictions of NASA images. These images show planets and moons that are similar to potential exoplanet environments, fueling their creativity for what their own exoplanet might look like.

\begin{table*}[t]
  \centering
  \begin{tabular}{|c|c|c|c|}
    \hline
    & Race or Ethnicity & \multicolumn{2}{|c|}{Number} \\
    \hline
    \multirow{4}{*}{URM} & Black or African American & 16 & \multirow{4}{*}{48} \\
    & Hispanic, Latinx or Spanish Origin & 18  & \\
    & American Indian or Alaska Native & 2 & \\
    & Mixed race URM & 12 & \\
    \hline
    \multirow{3}{*}{Non-URM} & Asian & 14 & \multirow{3}{*}{25}\\
    & Middle Eastern or North African & 2 & \\
    & White & 9 &  \\
    \hline
    Unspecified & Unspecified & 7 & 7\\
    \hline
    \hline
    \multicolumn{2}{|r|}{Total participants} & \multicolumn{2}{c|}{80}\\
    \hline
  \end{tabular}
  \vspace{0.5cm}
  \caption{Demographics of virtual workshop participants from Summer 2021, 2022, and 2023. Of the total, 48 identified as URM, 25 non-URM, and 7 preferred not to specify. The categories shown are based on the National Science Foundation categorizations \citep{NCSES:2023}.}
  \label{tab:demographics}
\end{table*}

\subsection{Participant Demographics}
\label{sec:demo}

The primary demographic of Rising Stargirls workshops are middle-school aged (11-14 years old). However, in order to welcome broad interest, occasional exceptions were made to accommodate participants who were younger or older than this range. Therefore, it is noted that some survey responses may correspond to students outside of the middle-school age range.

From 2015 to 2016, three in-person Rising Starigirls workshops were conducted by Professor Aomawa Shields with support from her NSF Astronomy and Astrophysics Postdoctoral Fellowship. 

In Winter 2015, Rising Stargirls collaborated with Irving STEAM Magnet Middle School in Eagle Rock, California to offer their students after-school programming. Irving STEAM aims to develop critical thinking by project-based learning that incorporates both the arts and sciences. In 2019, Irving STEAM's student population was 82\% Hispanic\footnote{\href{https://www.greatschools.org/california/los-angeles/2162-Irving-STEAM-Magnet/\#Students}{https://www.greatschools.org/california/los-angeles/2162-Irving-STEAM-Magnet/\#Students}}, and 100\% of students who participated in the Rising Stargirls after-school programming in 2015 self identified as Hispanic. This after-school programming took place for two hours twice a week for three weeks, for a total of 12 hours.

In Summer 2015, Rising Stargirls collaborated with the Science Club for Girls in Cambridge, Massachusetts for their intensive skill-development and career exploration program called "Young Leaders in STEM." Science Club for Girls offers experiential learning for girls and gender-expansive youth from underrepresented communities\footnote{\href{https://www.scienceclubforgirls.org/}{https://www.scienceclubforgirls.org/}}. Students who participated in the Rising Stargirls summer programming came from low-income backgrounds. This summer camp programming took place over four days within a week for four hours each day, for a total of 16 hours. 

In Fall 2016, Rising Stargirls collaborated with the Young Women's Christian Association (YWCA) in Pasadena-Foothill Valley. The YWCA's mission of "eliminating racism, empowering women, and promoting peace, justice, freedom and dignity for all," \footnote{\href{https://www.ywca.org/what-we-do/our-mission-in-action/}{https://www.ywca.org/what-we-do/our-mission-in-action/}} aligns with Rising Stargirls' mission of "encouraging girls of all colors and backgrounds to learn, explore, and discover the universe."\footnote{\href{http://www.risingstargirls.org/}{http://www.risingstargirls.org/}} In 2010, 34\% of Pasadena's population identified as Hispanic or Latino, and 10\% identified as Black. This after-school programming took place for two hours twice a week for three weeks, for a total of 12 hours.

With the onset of the COVID-19 pandemic in 2020, Rising Stargirls adapted its programming to a virtual format to continue reaching students during the isolation restraints. In Summer 2021, Rising Stargirls hosted its first virtual workshop. It reached 40 participants living in nine states across the United States. Of its participants, 57\% self identified with URM groups. This summer workshop ran for two hours a day, Monday through Friday for two weeks, for a total of 20 hours. 

In Summer 2022, Rising Stargirls hosted its second virtual workshop. It reached 28 participants, including four returning participants, from nine states across the United States. Of its participants, 69\% self identified with URM groups. This summer workshop ran for two hours a day, Monday through Friday for two weeks, for a total of 20 hours.

In Summer 2023, Rising Stargirls hosted its third virtual workshop. It reached 33 participants, including three returning participants, from 11 states across the United States and five countries outside of the United States. Of its participants, 48\% self identified with URM groups. This summer workshop ran for two hours a day, Monday through Friday for two weeks, for a total of 20 hours.

The demographics of the participants in the three virtual workshops are presented in Table~\ref{tab:demographics}. Demographics of participants in the in-person workshops were not available, so they are not included in Table~\ref{tab:demographics}. Of the 66 participants who filled out a demographic survey, 60\% identified with URM groups, with 20\% being Black or African American, 23\% Hispanic, Latinx, or Spanish origin, 3\% American Indian or Alaska Native, and 15\% having multiple identities, including at least one URM. The remaining 31\% of the participants identified with non-URM groups, with 18\% being Asian, 3\% Middle Eastern or North African, and 11\% White. The remaining 9\% preferred not to specify. These categories for participant race and ethnicity are based on the categories for URM by the \citet{NCSES:2023}. This demonstrates that the Rising Stargirls program has achieved its goal of targeting girls from URM backgrounds while also reaching a diverse group of girls and welcoming everyone who expresses interest in the workshops.

\subsection{Resources}
\label{sec:resources}

Rising Stargirls has made two main resources publicly available to encourage the integration of arts-based activities in science curriculum outside of the Rising Stargirls programming: activity handbooks and educator webinars. Although these resources are not immediately applicable to the findings of the paper, they are described here to paint a more complete picture of the Rising Stargirls organization.

To encourage educators across the world to integrate creative arts-based strategies into scientific curricula in their own classrooms, Rising Stargirls publicly released \textit{The Rising Stargirls Teaching and Activity Handbook}\footnote{You can access the handbook after filling out a short questionnaire at \href{http://www.risingstargirls.org/teaching-and-activity-handbook-1}{http://www.risingstargirls.org/teaching-and-activity-handbook-1}}, written by Professor Shields. At the onset of the COVID-19 pandemic, this handbook was adapted by Maya Silverman to allow curriculum activities to be conducted via Zoom and similar online platforms\footnote{You can access the Zoom adapted handbook after filling out a short questionnaire at \href{http://www.risingstargirls.org/covid-19-virtual-teaching-and-activity-handbook}{http://www.risingstargirls.org/covid-19-virtual-teaching-and-activity-handbook}}. In 2021, Jessica N. Howard developed five additional Rising Stargirls activities that can be conducted both virtually and in person. These activities are available in \textit{The Rising Stargirls Teaching Activity Supplement}\footnote{You can access the supplemental activities after filling out a short questionnaire at \href{http://www.risingstargirls.org/teaching-and-activity-supplement}{http://www.risingstargirls.org/teaching-and-activity-supplement}}. In 2021, Nina Robbins Blanch translated the \textit{Rising Stargirls Teaching and Activity Handbook} to Spanish, now publicly available on the Rising Stargirls website\footnote{You can access the Spanish language handbook after filling out a short questionnaire at \href{http://www.risingstargirls.org/teaching-and-activity-handbook-en-espanol}{http://www.risingstargirls.org/teaching-and-activity-handbook-en-espanol}}.

These free handbooks lead educators through over 20 hours of arts-based astronomy activities developed by Rising Stargirls and borrowed with permission from other programs\footnote{\href{https://dbp.theatredance.utexas.edu/node/29}{Drama-Based Instruction (DBI) Network}, \href{https://dbp.theatredance.utexas.edu}{Drama For Schools}, \href{ https://eurekus.org}{Eurekus}, \href{http://improvencyclopedia.org}{Improv Encyclopedia}, \href{https://astrobiology.nasa.gov/nai/media/medialibrary/2013/10/Astrobiology-Educator-Guide-2007.pdf}{NASA Astrobiology Institute} "Life on Earth...and elsewhere?" Astrobiology Teaching Guide, \href{https://www.nasa.gov/planetarymissions/discovery.html}{NASA's Discovery and New Frontiers} Programs, and \href{http://journeythroughtheuniverse.org/downloads/Content/Voyage!.pdf}{Voyage: A Journey Through Our Solar System.}}. The original handbook guides educators through 10 days of working with middle-school girls including ice breakers and an end-of-program celebration. The handbook supplement leads educators through five additional creative arts-based astronomy activities. Each activity  introduces a topic in astronomy or astrobiology and encourages exploration of the topic through a creative arts lens. 
 
Since 2019, Rising Stargirls has hosted five webinars geared towards modeling how educators can incorporate the Rising Stargirls philosophy into their own outreach programs. Participating educators ranged from those growing their program to those with an established presence in the communities they serve. These two hour webinars complemented the publicly available \textit{The Rising Stargirls Teaching and Activity Handbook}, as educators are guided by Rising Stargirls’ facilitators through a selection of learner-centered activities included in the handbook, and taught how to develop these activities for in-person and virtual classroom environments. In addition, educators received access to further resources beyond \textit{The Rising Stargirls Teaching and Activity Handbook}, including example workshop flyers, participant contracts, media and art consent and release forms, and directions to potential funding resources. These resources were distributed with the intent to be built upon and adapted by individual educators for the needs of their own outreach communities. In this way, the Rising Stargirls’ philosophy extends beyond the direct endeavors of Rising Stargirls, and intersects with underserved communities alongside their target group of URM. 

\section{Methodology}
\label{sec:methodology}

Data for this work were collected using anonymous surveys to gauge participants feelings before and after attending a Rising Stargirls workshop. The next two subsections describe how data were collected and analyzed to produce the results in Section~\ref{sec:results}. 

\subsection{Data Collection}
\label{sec:datacollec}

The data discussed in this section were originally collected as part of anonymous, program-improvement surveys conducted by Rising Stargirls between 2015 and 2023. We discuss the collection strategies used for this purpose. Data was collected voluntarily with no adverse consequences for participants who chose not to fill out the survey. All subsequent analysis was done on de-identified data. 

During the three in-person workshops, surveys were administered audibly and collected using a physical piece of paper. Data were compiled and saved to a spreadsheet. The same survey, detailed below, was administered on the first and last day of the workshop. During the virtual workshop in the summer of 2021, the initial survey was administered visually via slides at the beginning of the second day of the workshop. For the initial survey, no survey results were collected but the participants were asked to write their answers in their notes. The final survey was administered via Google Form at the end of the last day of the workshop. Responses for how participants felt before and after the workshop were collected via a Google Form on the last day of the workshop. During the virtual workshops in the summers of 2022 and 2023, the initial survey was administered via Google Form at the end of the first day of the workshop. For the initial survey, survey results were collected but not used for research purposes. The final survey was administered via Google Form at the end of the last day of the workshop. Responses for how participants felt before and after the workshop were collected via a Google Form on the last day of the workshop. 

The survey items comprised four statements:
\begin{enumerate}
    \item I talk to my family and friends about science.
    \item I like my science classes. (Altered to: I am excited to take my science classes in school.)\footnote{Statement 2 was changed for the virtual workshops to better reflect participants' attitudes towards future science classes.}
    \item I believe I can do well in science.
    \item I see myself as a scientist.		
\end{enumerate}

Participants were asked to rate each statement on a Likert scale of one to six. For the first statement, the scale read: Never (1), Very Rarely (2), Rarely (3), Occasionally (4), Frequently (5), Very Frequently (6). For the second, third, and fourth statements, the scale read: Disagree Very Strongly (1), Disagree Strongly (2), Disagree (3), Agree (4), Agree Strongly (5), Agree Very Strongly (6).

Evaluation items were developed in consultation with the Science Education Department at the Harvard-Smithsonian Center for Astrophysics. They were based on the \textit{Partnerships in Education and Resilience} Common Instrument Suite\footnote{\href{https://www.pearinc.org/common-instrument-suite}{https://www.pearinc.org/common-instrument-suite}} \citep{Noam2020TheCI} and the Persistence Research in Science and Engineering project \citep{Hazari2013TheSI}. Survey items one and two were developed to assess participants' engagement in science, while items three and four were developed to assess participants' science identity.

While the workshops conducted were focused on astronomy education, we study participants engagement and identity in science more broadly. Formal astronomy classes are often not offered in middle schools. This makes it difficult to ask participants about their experience in astronomy classes. Furthermore, our hypothesis applies across the sciences and is not confined to astronomy. Therefore, the assessment questions were intentionally designed to evaluate participants' engagement with science and science identity.

Data from the in-person workshops were saved as the number or percentage of participants who answered a question with each response. For example, at the beginning of one of the workshops, 50\% of participants stated that they rarely talked to their families and friends about science. In Winter 2015, 10 participants filled out the before survey and about 12 filled out the after survey. In Summer 2015, seven participants filled out the before and after surveys for question 1, six participants filled out the before survey for questions 2, 3 and 4, seven participants filled out the after survey for questions 2 and 3, and nine participants filled out the after survey for question 4. In Fall 2016, 9 participants filled out the before survey and 5 participants filled out the after survey. Inconsistencies in the number of participants who filled out the survey is due to the change in number of participants who attended on the first and last day of the workshops. 

The data-collection strategies used in the virtual workshops allowed for more details in saved data since surveys were administered with a Google Form. Data from these workshops were recorded verbatim in the manner each participant responded to respective survey questions before and after the workshop. For example, one participant stated that they talked to their families and friends about science occasionally (4) before the workshop and frequently (5) after the workshop. The 2022 and 2023 virtual workshops included several returning participants (see Section~\ref{sec:demo}). Their responses may be included in the data. In Summer 2021, 30 participants filled out the before and after survey. In Summer 2022, 15 participants filled out the before and after survey. In Summer 2023, 21 participants filled out the before and after survey.

\begin{table*}
    \centering
    \begin{tabular}{|l|l|l|}
        \hline
        \multicolumn{1}{|c|}{Summer 2021} & \multicolumn{1}{c|}{Summer 2022} & \multicolumn{1}{c|}{Summer 2023} \\
        \hline
        Draw a scientist & Draw a scientist & Draw a scientist \\
        When I look at the sky I think of... & When I look at the sky I think of... &  When I look at the sky I think of...\\
        Freewriting on Astro(biology) & Freewriting on Astro(biology) & Freewriting on Astro(biology) \\
        Why does representation matter? & Why does representation matter? & Why does representation matter? \\
        Constellations and origin stories & Constellations and origin stories & Constellations and origin stories \\
        Art and the Cosmic Connection & Art and the Cosmic Connection & Art and the Cosmic Connection \\
        Planet mnemonics & Telescopes and Kaleidoscopes & Telescopes and Kaleidoscopes \\
        Share a woman of color astronomer & Share a woman of color astronomer & Share a woman of color astronomer \\
        Drawing distance in me & Dark matter and hidden messages & Galaxies and their diversity \\
        Create your very own exoplanet & Create your very own exoplanet & Create your very own exoplanet \\
        Life's must-haves & Life's must-haves & Life's must-haves \\
        PSA for life  & PSA for life & PSA for life \\
        My universe & My universe & The expansion of the universe \\
        \hline

    \end{tabular}
    \caption{Titles of activities taught in each virtual workshop. The details of the corresponding activities can be found in the Rising Stargirls handbooks discussed in Section~\ref{sec:resources}.}
    \label{tab:activities}
\end{table*}

\subsection{Analysis}
\label{sec:analysis}

The data was analyzed using a Python (version 3.10.12) script that read data from CSV files. The goal of our analysis is to report how participants' engagement in and self-identity towards science changed over the duration of the workshop. Section~\ref{sec:results} presents figures that show how participants responded to the four statements. We also discuss how these observed changes support our claims that Rising Stargirls workshops improve engagement with science and science identity among participants. 

Data is analyzed cumulatively: data from all three in-person workshops are shown together and data from all three virtual workshops are shown together. Our results should not be dependent on the details that varied from workshop to workshop since they were run with similar arts-based activities (see Table~\ref{tab:activities} for details). However, we also analyze each workshop individually to comment on the general trend from workshop to workshop. 

To quantitatively analyze the data we convert the Likert scale responses (1 to 6) to a numeric scale (0 to 5),%
\footnote{Given that the Likert scale is fundamentally a set of discrete ordered categories, and not a continuous variable, it technically violates assumptions used in standard parametric statistical metrics. Thus, nonparametric statistical analyses are necessary, although they are generally considered less powerful~\citep{BishopHerron2015}. As a result, in this work, we report both parametric (equations~\ref{eq:HakesGain} and~\ref{eq:normChange}) and nonparametric metrics (equation~\ref{eq:dA}).}
and then convert these scores to percentages e.g. "Very Frequently" corresponds to a numeric score of 5 which is converted to 100\% while "Never" corresponds to a numeric score of 0 which is converted to 0\%.

We consider three different metrics to compare participants' responses before and after the workshops: Gain of averages (Hake’s Gain), $\langle g \rangle$~\citep{Hake1998}, normalized change, $c$~\citep{MarxCummings2007}, and a nonparametric effect size measure, $d_{A_w}$~\citep{Li2016-fy, Ruscio2008-vl}.

The gain of averages (Hake’s Gain), $\langle g \rangle$~\citep{Hake1998}, is defined as
\begin{equation} \label{eq:HakesGain}
    \langle g \rangle \equiv \frac{A_{\rm{ave}} - B_{\rm{ave}}}{100\% - B_{\rm{ave}}},
\end{equation}
where $A_{\rm{ave}}$ is the average percent response after the workshop and $B_{\rm{ave}}$ is the average percent response before the workshop.

Normalized change, $c$~\citep{MarxCummings2007}, is similar to Hake's gain but has several beneficial features such as being able to account for score losses and being more robust for small class sizes. However, it is limited in that it requires matched data (i.e. that we can match a participant's responses before the workshop to their responses after). It is defined as

\begin{equation} \label{eq:normChange}
    c \equiv 
    \begin{cases}
    \frac{r_{{\rm A, i}} - r_{{\rm B, i}}}{100 \% - r_{{\rm B, i}}} & r_{{\rm A, i}} > r_{{\rm B, i}}\\
    \text{dropped~from~calculation} & r_{{\rm A, i}} = r_{{\rm B, i}} = 100 \% \text{~or~} 0 \%\\
    0 & r_{{\rm A, i}} = r_{{\rm B, i}}\\
    \frac{r_{{\rm A, i}} - r_{{\rm B, i}}}{r_{{\rm B, i}}} & r_{{\rm A, i}} < r_{{\rm B, i}}\\
    \end{cases},
\end{equation}
where $r_{\rm{A,i}}$ ($r_{\rm{B,i}}$) is the percent score after (before) the workshop of participant $\rm{i}$. $c$ is then averaged over all participants which were not dropped from the calculation to get $c_{ave}$.

Since our data is non-normal, we estimate our effect size with the nonparametric effect size measure, $d_{A_w}$~\citep{Li2016-fy}, derived from the nonparametric estimator for CL, $A_w$~\citep{Ruscio2008-vl}, which is defined as
\begin{equation}
    A_w \equiv \frac{ \#\left( r_{\rm A} > r_{\rm B} \right) + 0.5 \#\left( r_{\rm A} = r_{\rm B} \right) }{N_{\rm A} N_{\rm B}},
\end{equation}
where $N_{\rm A}$ ($N_{\rm B}$) is the number of responses after (before) the workshop and $\# \left( \cdot \right)$ is the count function.  Namely, for each after response, $r_{\rm A, i}$, $\#\left( r_{\rm A, i} > r_{\rm B} \right)$ counts the number of before responses which are less than $r_{\rm A, i}$. To better interpret this result, $A_w$ can be converted into a $d-$metric via
\begin{equation} \label{eq:dA}
    d_{A_w} \equiv \sqrt{2} \Phi^{-1}\left(A_w\right),
\end{equation}
where $\Phi^{-1}\left(\cdot\right)$ is the inverse normal cumulative distribution function~\citep{Li2016-fy}.

We calculate $\langle g \rangle$ and $d_{A_w}$ for all workshop data but can only calculate $c_{ave}$ on the matched virtual workshop data. The in-person data was saved cumulatively and anonymously, as described in Section~\ref{sec:datacollec}, so it was impossible to link a participants' responses before the workshop to those after the workshop. Because it is unmatched, calculating $c_{ave}$ on the in-person workshop data is not possible.

\section{Results}
\label{sec:results}
In Sections~\ref{sec:IP} and~\ref{sec:O} we discuss the results using data from Rising Stargirls' in-person and virtual workshops, respectively. The data are split between in-person and virtual workshops to study each method individually. Each method was studied individually to investigate the impact of slight differences in instruction media and curriculum between workshops, which may have slightly impacted interactions between participants. Furthermore, the collection method differed between the two workshop types, allowing for different analyses, as described in Section~\ref{sec:analysis}.

\begin{table*}
    \centering
    
    \begin{tabular}{l|m{2.5cm}||c|c|c||c}
        Assessment Item & \centering Metric with Number of Responses & Winter 2015 & Summer 2015 & Fall 2016 & Total \\
        \hline
        \multirow{3}{*}{I talk to my family and friends about science.}  & \centering ($N_B$, $N_A$) & (10, 12) & (7, 7) & (9, 5) & (26, 24)\\
        & \centering  $d_{A_w}$               & 0.2524 & 0.7583 & 0.8336 & 0.4268 \\
        & \centering $\langle g \rangle$ & 0.0885 & 0.5000 & 0.2909 & 0.2344 \\
        \hline
        \multirow{3}{*}{I like my science classes.}  & \centering ($N_B$, $N_A$) & (10, 12) & (6, 7) & (9, 5) & (25, 24)\\
        & \centering $d_{A_w}$               & 0.3279 & 0.1268 & 0.8336 & 0.4258 \\
        & \centering $\langle g \rangle$ & 0.3561 & 0.2381 & 0.7000 & 0.4048 \\
        \hline
        \multirow{3}{*}{I believe I can do well in science.}  & \centering ($N_B$, $N_A$) & (10, 13) & (6, 7) & (9, 5) & (25, 25)\\
        & \centering $d_{A_w}$               & -0.2744 & -0.2546 & 0.5238 & -0.0255 \\
        & \centering $\langle g \rangle$ & -0.3077 & -0.7143 & 0.3571 & -0.0714 \\
        \hline
        \multirow{3}{*}{I see myself as a science person.}  & \centering ($N_B$, $N_A$) & (10, 12) & (6, 9) & (9, 5) & (25, 26)\\
        & \centering $d_{A_w}$               & 0.9354 & -0.9955 & 0.8336 & 0.4463 \\
        & \centering $\langle g \rangle$ & 0.3382 & -2.1110 & 0.2909 & 0.2582 \\
        \hline
    \end{tabular}
    \caption{Statistical analysis of in-person workshop responses. Since these data are unmatched we report two statistical metrics: $d_{A_w}$ (equation~\ref{eq:dA}) and $\langle g \rangle$ (equation~\ref{eq:HakesGain}). The number of before, $N_B$, and after, $N_A$, responses used in these calculations are also reported. The sign of the value indicates a gain (+) or loss (-). 
    A small effect is indicated by $|\langle g \rangle|<0.30$ or $|d_{A_w}| \sim 0.20$.
    A moderate effect is indicated by $0.30 \leq |\langle g \rangle|<0.70$ or $|d_{A_w}| \sim 0.50$.
    A large effect is indicated by $0.70 \leq |\langle g \rangle|$ or $|d_{A_w}| \sim 0.80$.
    }
    \label{tab:inPerson}
\end{table*}

\subsection{In-Person}
\label{sec:IP}

\begin{figure*}
    \centering
    \begin{subfigure}[b]{0.45\textwidth}
         \centering
         \includegraphics[width=\textwidth]{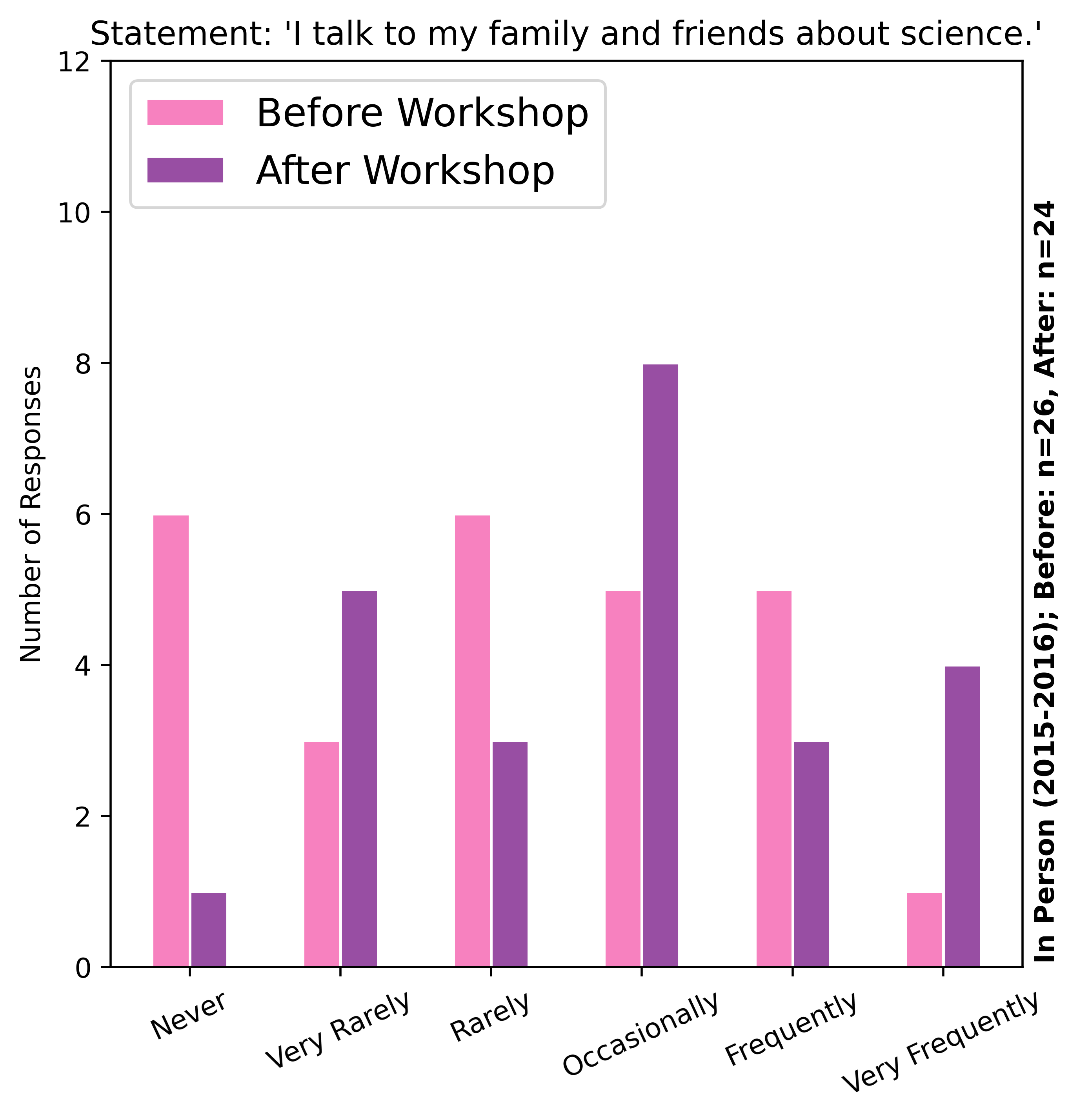}
         \caption{Responses to the statement “I talk to my family and friends about science.” Responses range from “Never” with a numerical value of 0 (or 0\%) to “Very Frequently” with a numerical value of 5 (or 100\%).}
         \label{fig:talkaboutscienceIP}
     \end{subfigure}
     \hfill
     \begin{subfigure}[b]{0.45\textwidth}
         \centering
         \includegraphics[width=\textwidth]{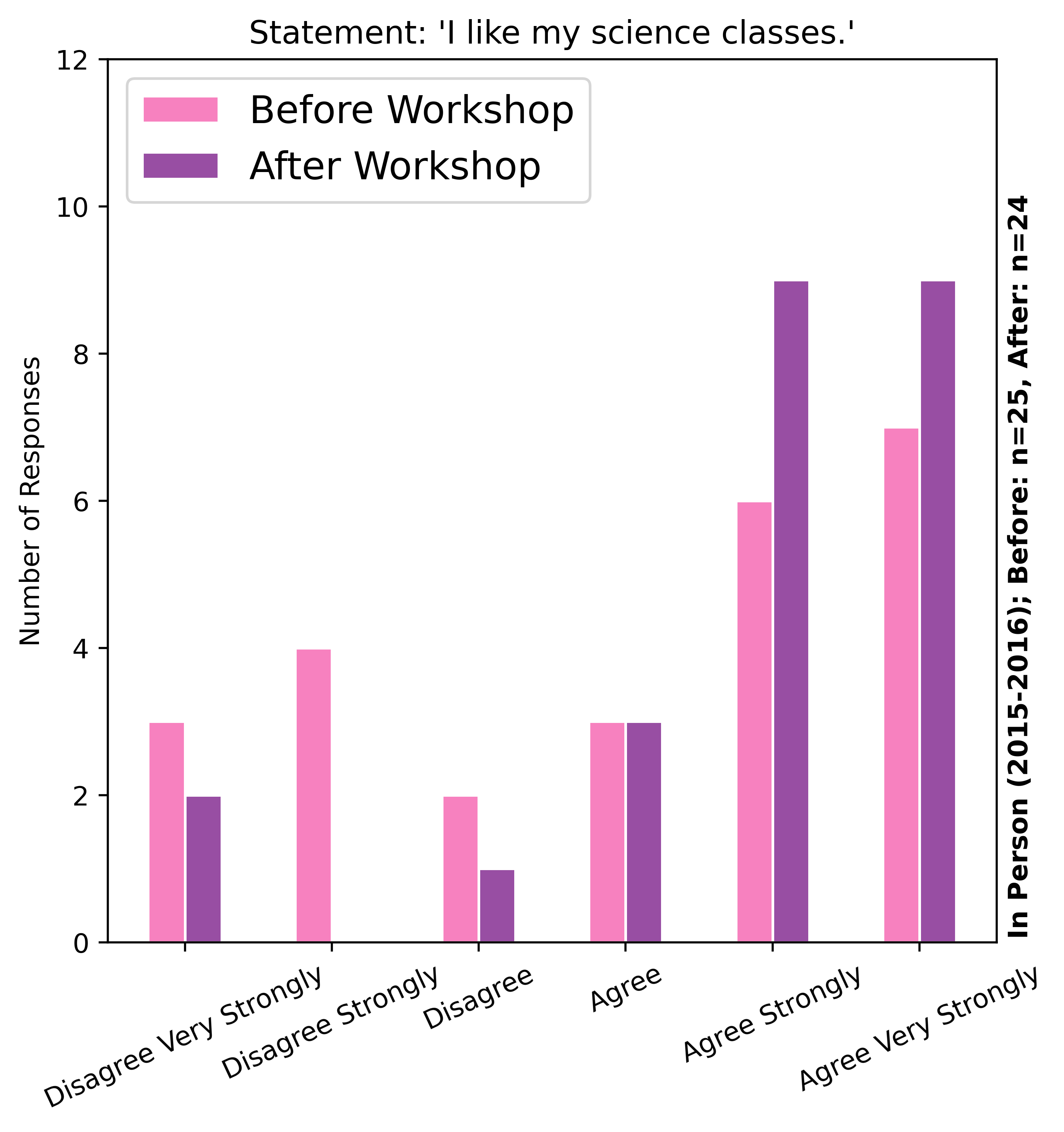}
         \caption{Responses to the statement “I like my science classes.” Responses range from “Disagree very strongly” with a numerical value of 0 (or 0\%) to “Agree very strongly” with a numerical value of 5 (or 100\%).}
         \label{fig:likeclassesIP}
     \end{subfigure}
     \hfill
    \begin{subfigure}[b]{0.45\textwidth}
         \centering
         \includegraphics[width=\textwidth]{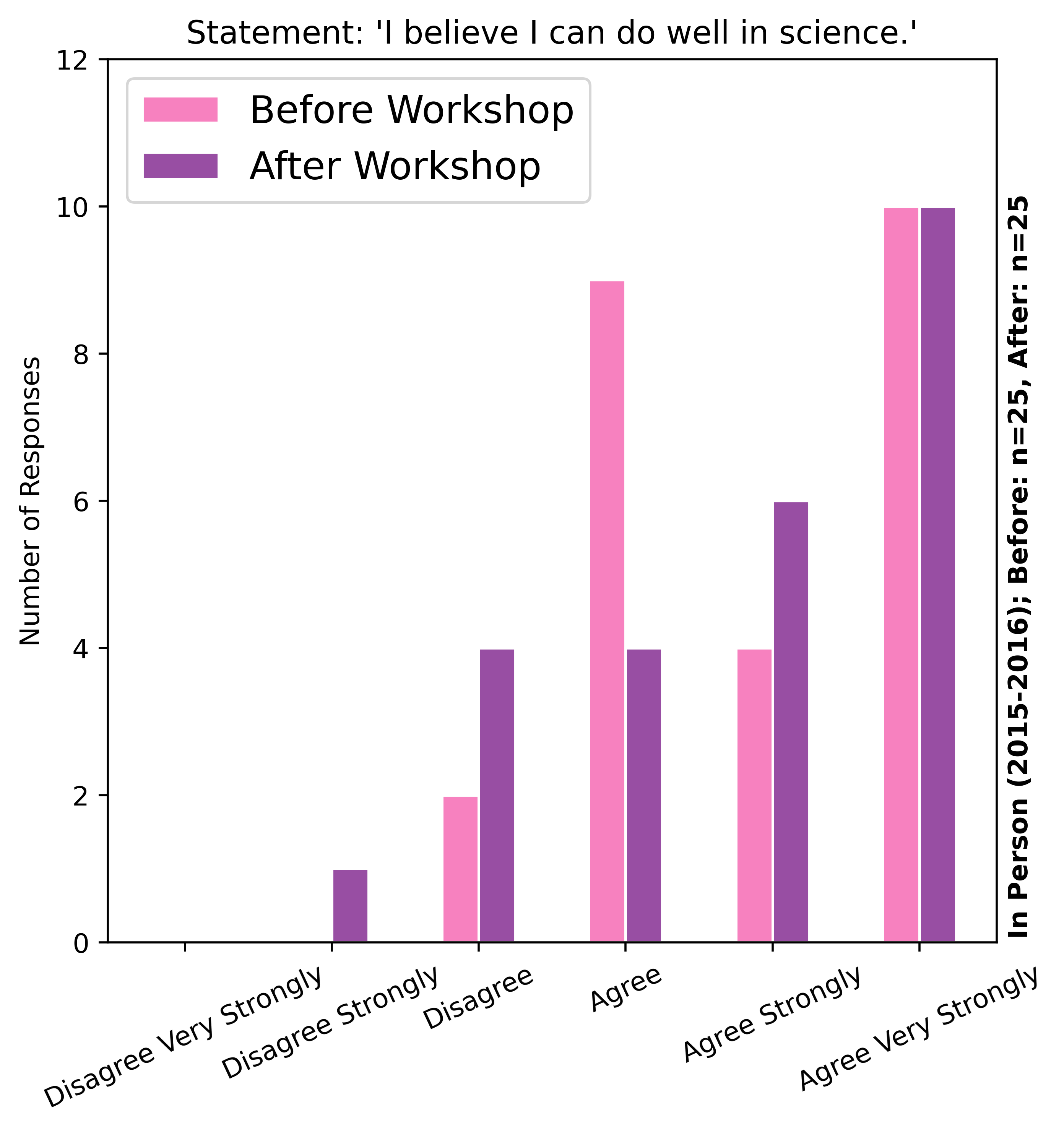}
         \caption{Responses to the statement “I believe I can do well in science.” Responses range from “Disagree very strongly” with a numerical value of 0 (or 0\%) to “Agree very strongly” with a numerical value of 5 (or 100\%).}
         \label{fig:dowellinscienceIP}
     \end{subfigure}
     \hfill
     \begin{subfigure}[b]{0.45\textwidth}
         \centering
         \includegraphics[width=\textwidth]{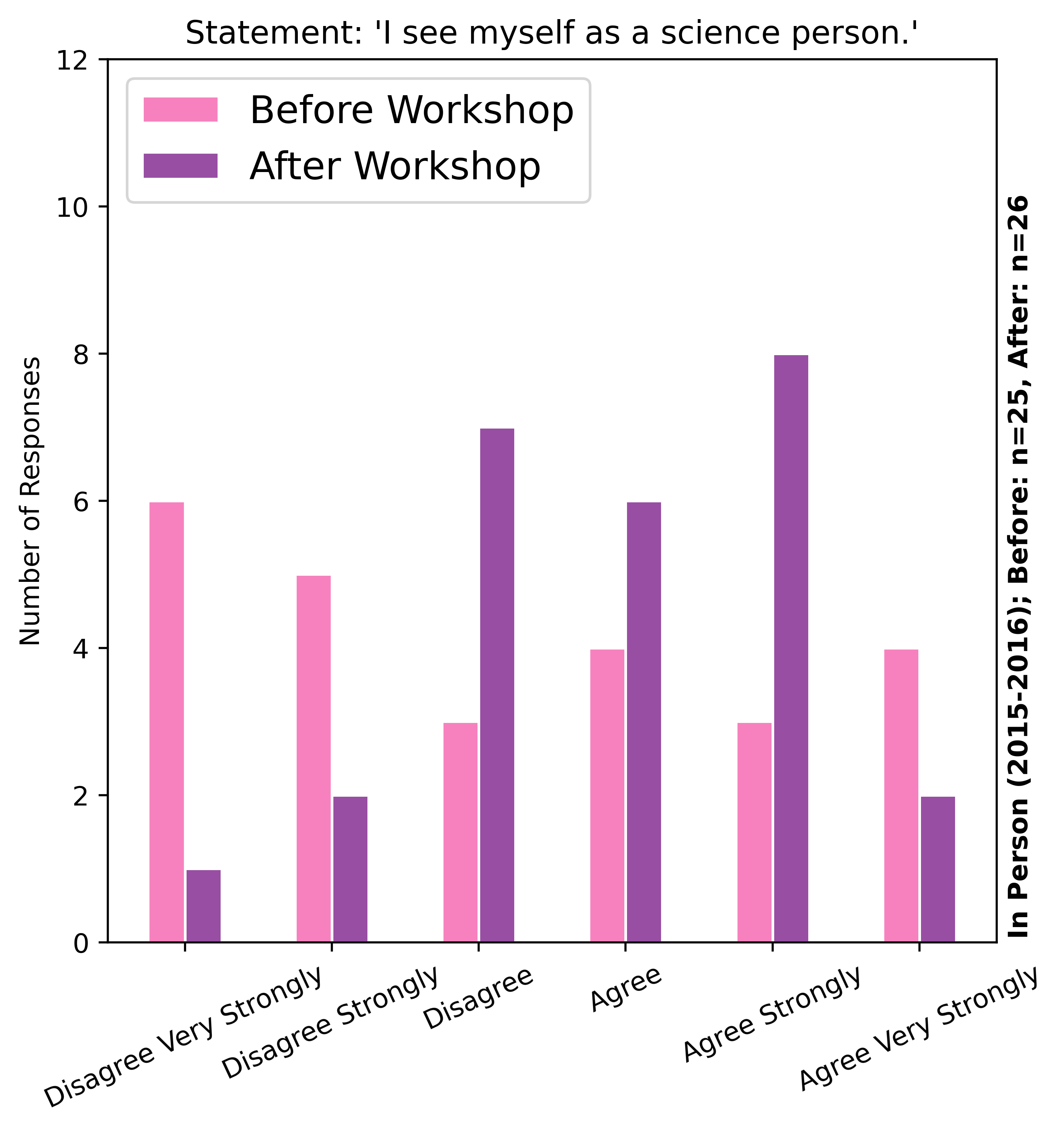}
         \caption{Responses to the statement “I see myself as a science person.” Responses range from “Disagree very strongly” with a numerical value of 0 (or 0\%) to “Agree very strongly” with a numerical value of 5 (or 100\%).}
         \label{fig:sciencepersonIP}
     \end{subfigure}
    \caption{Compiled data from three in-person workshops that took place in Winter 2015, Summer 2015, and Fall 2016. The y-axis shows how many participants responded to the statement for each response. Responses from before the workshop are shown in pink and are on the left side of each response, and responses from after the workshop are shown in purple and are on the right side. }
    \label{fig:allIP}
\end{figure*}

Figure~\ref{fig:talkaboutscienceIP} demonstrates that, overall, participants engaged in more frequent discussions about science with their families and friends after participating in one of Rising Stargirls' in-person workshops. Responses from before the workshop are shown in pink and are on the left side of each response, and responses from after the workshop are shown in purple and are on the right side. We assigned numbers to participant responses, such that 0 (or 0\%) corresponds to "Never" and 5 (or 100\%) corresponds to "Very Frequently". We then compare before and after responses by calculating Hake's gain, $\langle g \rangle$, and the effect size, $d_{A_w}$,  which are reported in Table~\ref{tab:inPerson} (see Section~\ref{sec:analysis} for definitions). 
Looking at each workshop individually, there was a small to moderate increase in the frequency with which participants talked to their families and friends about science. Collectively analyzing all in-person workshop responses $(N_B, N_A) = (26, 24)$ showed a small gain $\langle g \rangle = 0.2344$ and small-to-moderate effect size of $d_{A_w} = 0.4268$.

Figure~\ref{fig:likeclassesIP} demonstrates that, overall, participants stated that they liked their science classes with stronger agreement after participating in one of Rising Stargirls' in-person workshops. We assigned numbers to participant responses, such that 0 (or 0\%) corresponds to "Disagree Very Strongly" and 5 (or 100\%) corresponds to "Agree Very Strongly". We then compare before and after responses by calculating Hake's gain, $\langle g \rangle$, and the effect size, $d_{A_w}$,  which are reported in Table~\ref{tab:inPerson} (see Section~\ref{sec:analysis} for definitions). 
Looking at each workshop individually, there was a small to moderate increase in the degree to which participants liked their science classes. Collectively analyzing all in-person workshop responses $(N_B, N_A) = (25, 24)$ showed a moderate gain $\langle g \rangle = 0.4048$ and small-to-moderate effect size of $d_{A_w} = 0.4258$.

\textit{The increase in the frequency with which participants talked to their families and friends about science, as well as the increase in the agreement that they like their science classes, as shown in Figures~\ref{fig:talkaboutscienceIP} and~\ref{fig:likeclassesIP}, indicates that participants' engagement in science increased over the course of Rising Stargirls' creative arts-based astronomy workshop.}

Figure~\ref{fig:dowellinscienceIP} demonstrates that, overall, participant's feelings of how well they believe they can do in science didn't change by a significant amount.  
Collectively analyzing all in-person workshop responses $(N_B, N_A) = (25, 25)$ showed a very slight decrease $\langle g \rangle = -0.0714$ and a very small effect size of $d_{A_w} = -0.0255$.
While this is a slight decrease we note that this assessment item received the highest rated results of all items both before and after the in-person workshops.
It is likely that their confidence in their ability to do well in science didn't increase because they came to the workshop already confident.
Interestingly, the trend changed drastically between each individual workshop, ranging from large to moderate decreases to moderate increase. 
We note that the most recent in-person workshop, Fall 2016, had a moderate increase, potentially indicating that participants benefited from improved program implementation as the program facilitators gained experience.

Figure~\ref{fig:sciencepersonIP} demonstrates that, overall, participants saw themselves as a science person with stronger agreement after participating in one of Rising Stargirls' in-person workshops. 
Collectively analyzing all in-person workshop responses $(N_B, N_A) = (25, 26)$ showed a small increase $\langle g \rangle = 0.2582$ and a small-to-moderate effect size of $d_{A_w} = 0.4463$. Individually, this trend also changed somewhat between workshops. Both the Winter 2015 and Fall 2016 workshops aligned with the collective result, showing moderate to small increases, respectively. 
However, the Summer 2015 workshop strongly deviated, showing a large decrease. We note that this deviation may be partially due to the fact that this workshop had 3 more responses for the after survey ($(N_B, N_A) = (6, 9)$).

\textit{Overall, the increase in agreement that participants saw themselves as a science person, shown in Figure~\ref{fig:sciencepersonIP}, indicates that participants positively developed their science identities over the course of Rising Stargirls' creative arts-based astronomy workshop.}

\subsection{Virtual}
\label{sec:O}

\begin{figure*}
    \centering
     \begin{subfigure}[b]{0.45\textwidth}
         \centering
         \includegraphics[width=\textwidth]{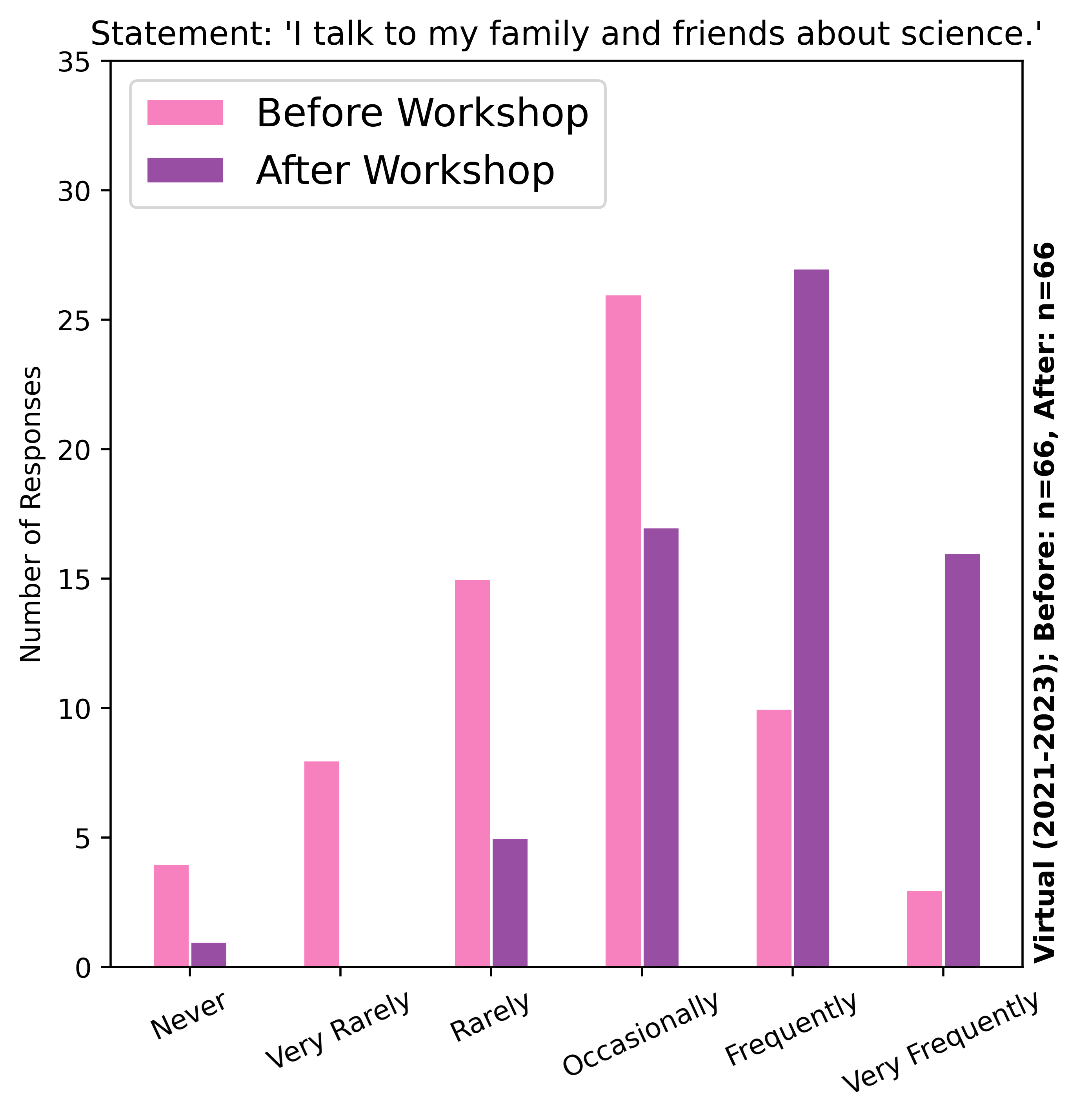}
         \caption{Responses to the statement “I talk to my family and friends about science.” Responses range from “Never” with a numerical value of 0 (or 0\%) to “Very Frequently” with a numerical value of 5 (or 100\%).}
         \label{fig:talkaboutscienceO}
     \end{subfigure}
     \hfill
     \begin{subfigure}[b]{0.45\textwidth}
         \centering
         \includegraphics[width=\textwidth]{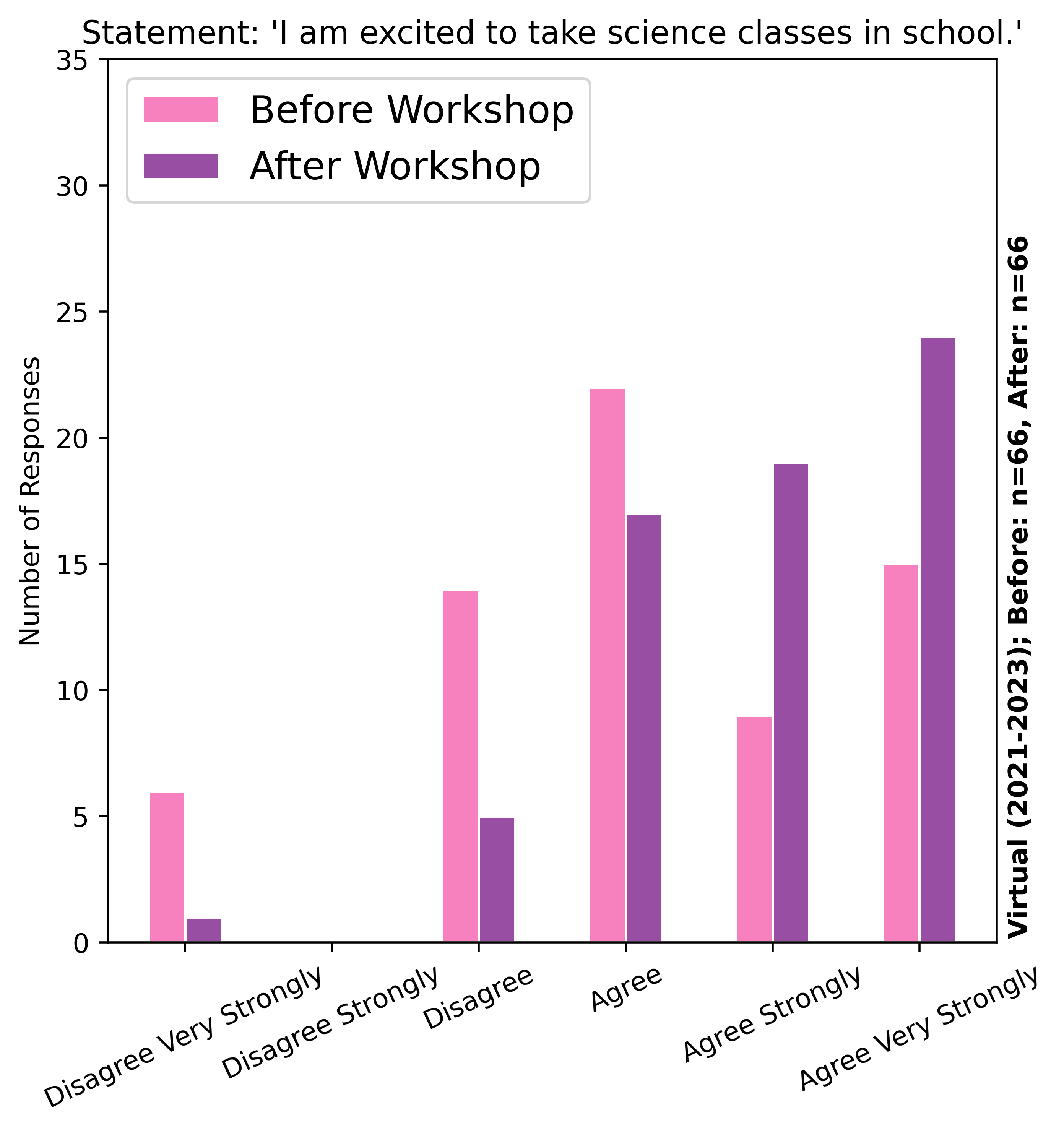}
         \caption{Responses to the statement “I am excited to take science classes in school.” Responses range from “Disagree very strongly” with a numerical value of 0 (or 0\%) to “Agree very strongly” with a numerical value of 5 (or 100\%).}
         \label{fig:excitedforscienceO}
     \end{subfigure}
     \hfill
     \begin{subfigure}[b]{0.45\textwidth}
         \centering
         \includegraphics[width=\textwidth]{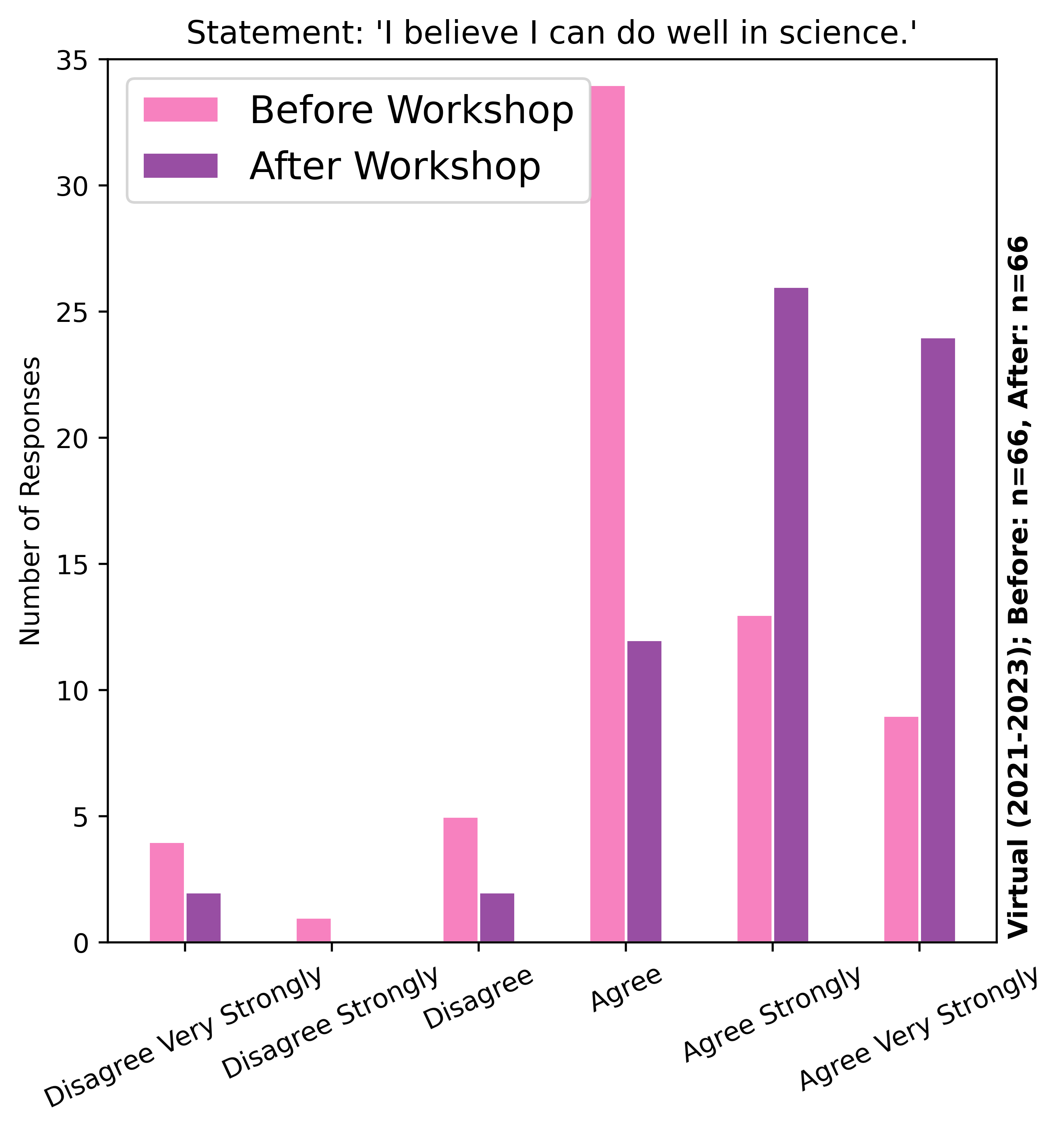}
         \caption{Responses to the statement “I believe I can do well in science.” Responses range from “Disagree very strongly” with a numerical value of 0 (or 0\%) to “Agree very strongly” with a numerical value of 5 (or 100\%).}
         \label{fig:dowellinscienceO}
     \end{subfigure}
     \hfill
     \begin{subfigure}[b]{0.45\textwidth}
         \centering
         \includegraphics[width=\textwidth]{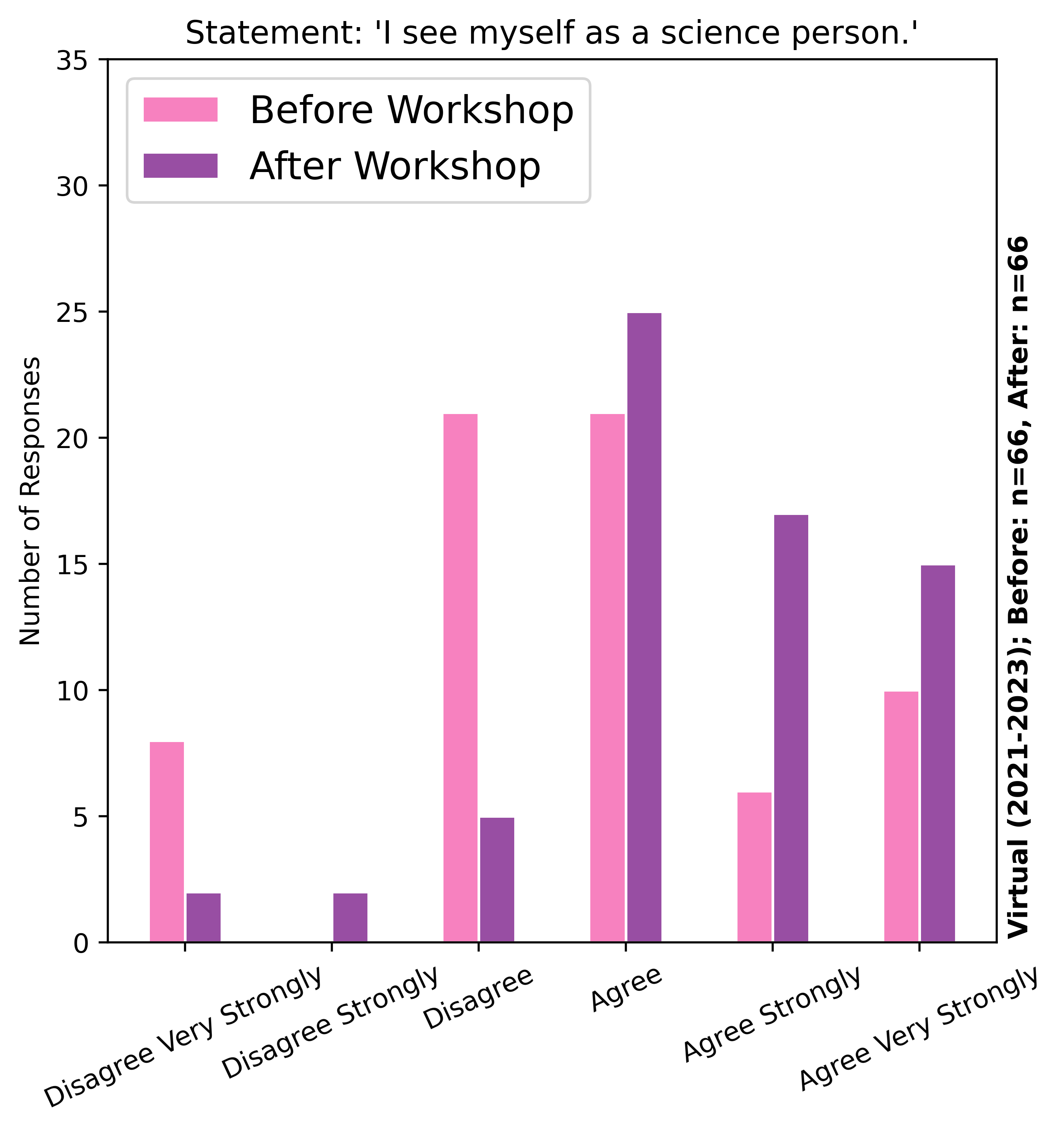}
         \caption{Responses to the statement “I see myself as a science person.” Responses range from “Disagree very strongly” with a numerical value of 0 (or 0\%) to “Agree very strongly” with a numerical value of 5 (or 100\%).}
         \label{fig:sciencepersonO}
     \end{subfigure}
    \caption{Compiled data from three virtual workshops that took place in Summer 2021, Summer 2022, and Summer 2023. The y-axis shows how many participants responded to the statement for each response. Responses from before the workshop are shown in pink and are on the left side of each response, and responses from after the workshop are shown in purple and are on the right side.}
\end{figure*}

\begin{table*}
    \centering
    
    \begin{tabular}{l|m{2.5cm}||c|c|c||c}
        Assessment Item & \centering Metric with Number of Responses & Summer '21 & Summer '22 & Summer '23 & Total \\
        \hline
        \multirow{3}{*}{I talk to my family and friends about science.}  & \centering ($N_B$, $N_A$) & (30, 30) & (15, 15) & (21, 21) & (66, 66)\\
        & \centering  $d_{A_w}$           & 1.3400 & 1.0500 & 0.8625 & 1.0700 \\
        & \centering $\langle g \rangle$  & 0.5846 & 0.4186 & 0.4314 & 0.4906 \\
        & \centering $c_{ave}$ (\# dropped) & 0.6034 (1) & 0.4189 (0) & 0.3921 (2) & 0.4958 (3) \\
        \hline
        \multirow{3}{*}{I am excited to take science classes in school.}  & \centering ($N_B$, $N_A$) & (30, 30) & (15, 15) & (21, 21) & (66, 66)\\
        & \centering $d_{A_w}$            & 0.8786 & 0.3909 & 0.4708 & 0.5900 \\
        & \centering $\langle g \rangle$  & 0.5385 & 0.2667 & 0.3721 & 0.4160 \\
        & \centering $c_{ave}$ (\# dropped) & 0.5417 (6) & 0.3091 (4) & 0.3222 (6) & 0.4247 (16) \\
        \hline
        \multirow{3}{*}{I believe I can do well in science.}  & \centering ($N_B$, $N_A$) & (30, 30) & (15, 15) & (21, 21) & (66, 66)\\
        & \centering $d_{A_w}$            & 1.1460 & 0.3016 & 0.9213 & 0.8201 \\
        & \centering $\langle g \rangle$  & 0.5472 & 0.2593 & 0.4500 & 0.4500 \\
        & \centering $c_{ave}$ (\# dropped) & 0.5617 (3) & 0.4875 (7) & 0.4450 (1) & 0.5085 (11) \\
        \hline
        \multirow{3}{*}{I see myself as a science person.}  & \centering ($N_B$, $N_A$) & (30, 30) & (15, 15) & (21, 21) & (66, 66)\\
        & \centering $d_{A_w}$            & 1.0710 & 0.3097 & 0.4285 & 0.6235 \\
        & \centering $\langle g \rangle$  & 0.4545 & 0.2105 & 0.2766 & 0.3377 \\
        & \centering $c_{ave}$ (\# dropped) & 0.5000 (4) & 0.2242 (4)	& 0.2314 (4) & 0.3593 (12) \\
        \hline
    \end{tabular}
    
    \caption{Statistical analysis of virtual workshop responses. Since these data are matched we report three statistical metrics: $d_{A_w}$ (equation~\ref{eq:dA}), $\langle g \rangle$ (equation~\ref{eq:HakesGain}), and $c_{ave}$ (equation~\ref{eq:normChange}). For $c_{ave}$, the number of responses dropped from the calculation are indicated in parentheses. The number of before, $N_B$, and after, $N_A$, responses used in these calculations are also reported. The sign of the value indicates a gain (+) or loss (-). 
    A small effect is indicated by $c_{ave}~\text{or}~|\langle g \rangle|<0.30$  or $|d_{A_w}| \sim 0.20$.
    A moderate effect is indicated by $0.30 \leq c_{ave}~\text{or}~|\langle g \rangle|<0.70$  or $|d_{A_w}| \sim 0.50$.
    A large effect is indicated by $0.70 \leq c_{ave}~\text{or}~|\langle g \rangle|$  or $|d_{A_w}| \sim 0.80$.
    }
    \label{tab:virtual}
\end{table*}

The virtual workshops varied slightly from the in-person workshops due to their virtual nature. Fewer theater exercises were included since many such activities demand face-to-face interactions. While virtual workshop participants were encouraged to use cameras to engage in face-to-face interactions, not all participants had access to technology allowing for face-to-face interactions. In addition to impacting the curriculum, the virtual nature also impacted the level of interactions between participants. In the in-person workshops, participants often knew each other already, and were able to interact with each other throughout the workshop. In the virtual workshops, participants did not know any other participants before the workshop, and interacted with each other via chat, voice, and occasionally with video. 
Additionally, as described in Section~\ref{sec:datacollec}, the data collection strategy changed for virtual workshops.
These differences may have impacted the following results.

To quantitatively analyze responses, we assigned numbers to each response such that 0 (or 0\%) corresponds to "Never" and 5 (or 100\%) corresponds to "Very Frequently". 
We then compare before and after responses by calculating Hake's gain, $\langle g \rangle$, and the effect size, $d_{A_w}$. Additionally, due to differences in collection strategies, the before and after virtual workshop responses are matched. This allows us to also calculate the normalized change, $c_{ave}$. These metrics are reported in Table~\ref{tab:virtual} (see Section~\ref{sec:analysis} for definitions). 

Figure~\ref{fig:talkaboutscienceO} demonstrates that, overall, participants engaged in more frequent discussions about science with their families and friends after taking part in one of Rising Stargirls' virtual workshops. 
Collectively analyzing all virtual workshop responses $(N_B, N_A) = (66, 66)$ to the question `I talk to my family and friends about science' showed a moderate increase $\langle g \rangle = 0.4906$ and $c_{ave} = 0.4958$ (with $3$ responses dropped from the calculation), as well as a large effect size $d_{A_w} = 1.0700$. The same trends hold for each individual virtual workshop.

Figure~\ref{fig:excitedforscienceO} demonstrates that, overall, participants were more excited to take science classes in school after participating in one of Rising Stargirls' virtual workshop. Collectively analyzing all virtual workshop responses $(N_B, N_A) = (66, 66)$ to the question `I am excited to take science classes in school' showed a moderate increase $\langle g \rangle = 0.4160$ and $c_{ave} = 0.4247$ (with $16$ responses dropped from the calculation), as well as a moderate-to-large effect size $d_{A_w} = 0.5900$. Each individual workshop shows similar trends with the Summer 2022 and Summer 2023 workshops showing slightly less increase across all metrics and the Summer 2021 workshop showing slightly more increase across all metrics.

\textit{The increase in the frequency with which participants talked to their families and friends about science, as well as the increase in their excitement to take science classes in school, as shown in Figures~\ref{fig:talkaboutscienceO} and~\ref{fig:excitedforscienceO}, indicates that participants' engagement with science increased over the course of Rising Stargirls' virtual creative arts-based astronomy workshop.}

Figure~\ref{fig:dowellinscienceO} demonstrates that, overall, participants believed that they could do well in science more strongly after participating in one of Rising Stargirls' virtual workshops. Collectively analyzing all virtual workshop responses $(N_B, N_A) = (66, 66)$ to the question `I believe I can do well in science' showed a moderate increase $\langle g \rangle = 0.4500$ and $c_{ave} = 0.5085$ (with $11$ responses dropped from the calculation), as well as a large effect size $d_{A_w} = 0.8201$. Individually, the Summer 2021 and Summer 2023 workshops show the same increase and the Summer 2022 workshop shows slightly less increase across all metrics.

Figure~\ref{fig:sciencepersonO} demonstrates that, overall, participants saw themselves as a science person with stronger agreement after participating in one of Rising Stargirls' virtual workshops. 
Collectively analyzing all virtual workshop responses $(N_B, N_A) = (66, 66)$ to the question `I see myself as a science person' showed a moderate increase $\langle g \rangle = 0.3377$ and $c_{ave} = 0.3593$ (with $12$ responses dropped from the calculation), as well as a large effect size $d_{A_w} = 0.6235$. Each individual workshop shows similar trends with the Summer 2022 and Summer 2023 workshops showing slightly less increase across all metrics and the Summer 2021 workshop showing slightly more increase across all metrics.

\textit{The increase in participants' belief in their capability to succeed in science, as well as the increase in agreement with which participants saw themselves as a science person, shown in Figures~\ref{fig:dowellinscienceO} and~\ref{fig:sciencepersonO}, indicates that participants positively developed their science identities over the course of Rising Stargirls' virtual creative arts-based astronomy workshop.}

We conducted an equivalent analysis on each of the virtual workshops individually to discuss any anomalies and trends. Table~\ref{tab:virtual}, shows that there is no clear trend from the earlier to later workshops across all four survey items on average. Rising Stargirls will continue tracking this trend for future workshops with the hope to produce more conclusive results on how the average percentage increase in participants' answers changes as the workshop progresses.

Additionally, for these workshops, since data was collected for each participant, we were able to track how a participant's response to each statement changed over the course of the workshop. Therefore, Table~\ref{tab:virtual} also shows the normalized change, $c_{ave}$, calculated using matched before and after responses from each participant for each individual workshop. In general, this largely agrees with the results of calculating Hake's gain, $\langle g \rangle$ on this dataset.

\section{Discussion}
\label{sec:discussion}

This study emphasizes the advantages of fostering creativity in informal science education for middle-school girls from underrepresented groups in the sciences. By analyzing participants' sentiments before and after taking part in a creative arts-based astronomy workshop run by Rising Stargirls, our results illustrate how encouraging middle-school girls to infuse the creative arts into science cultivates their science identity and encourages increased engagement with science. Science identity is assessed by asking participants whether they believe they can do well in science and if they see themselves as scientists. Engagement with science is evaluated by asking participants if they talk about science and if they enjoy their science classes.

Across all six workshops studied in this work, our results show that participants' engagement with science and science identity increased over the course of Rising Stargirls' creative arts-based astronomy workshops. 
Participants both talked to their families and friends about science more frequently and were more excited to take science classes in school after participating in one of Rising Stargirls' workshops. This indicates that participants' engagement with science increased over the course of Rising Stargirls' in-person and virtual workshops. Participants reported an increased belief in their ability to succeed in science and also in their identity as a science person. This indicates that participants' science identities increased over the course of Rising Stargirls' in-person and virtual workshops.

The workshop that took place in Summer 2015 has anomalously low gain and effect size. At one week long, this was Rising Stargirls' shortest workshop. We suspect this low improvement in confidence and enthusiasm towards their perceived science abilities is due to the lower relative amount of time available to change their opinions.
Alternatively, this could indicate that Rising Stargirls workshops improved over time and participants were positively impacted.
These could be reasons why it is deviating from the trend seen in other in person workshops. However, the low gain and effect size in this workshop does not change the overall trend seen in Figure~\ref{fig:allIP} and Table~\ref{tab:inPerson}.

\subsection{Limitations}

We acknowledge the limitations of retrospective pre- and post-surveys, where participants were asked to measure how they felt before and after the workshop on the final day. This depends on a participant’s memory recall, and the potential for participants to exhibit self enhancement bias \citep{PhysRevPhysEducRes.19.020167}. Alternatively, a pre-survey administered on the first day of the workshop, followed by a post-survey administered on the final day, would provide measurements not subject to inconsistencies within memory recall and allows educators to track participant engagement from the onset of the workshop \citep{PhysRevPhysEducRes.19.020167}. However, prior to the workshop, participants are likely to mischaracterize their comfort level with topics based on lack of exposure to overarching concepts \citep{PhysRevPhysEducRes.19.020167}. Once participants are introduced to the topics, in an active learning environment, they are more accurately able to describe the scope of what they did not know in comparison to what they have now learned \citep{PhysRevPhysEducRes.19.020167}. The workshop experience changes the metric by which participants assess themselves relative to what they thought they knew prior to the experience.

While our in-person workshops administered surveys on the first and last days of the workshop, ultimately, retrospective pre- and post-surveys were selected for the virtual workshops to decrease this response shift bias. The difference in survey administration over in-person and virtual workshops may lead to biases in our analysis and conclusions.

While our overall goal is to highlight the benefits of creative arts-based astronomy activities specifically for girls from URM groups, this study emphasizes the impacts on middle-school girls of \textit{all} colors and backgrounds. This choice was made to protect participant anonymity by ensuring that demographic information was not linked to survey responses. However, Rising Stargirls outreach efforts were designed to increase the participation of URM groups. As a result, URM individuals made up 60\% of the total workshop attendees. We acknowledge the value of understanding the impact on specific URM groups and will analyze this trend in future studies. This will allow us to explore which URM group benefited the most from creative arts-based astronomy workshops.

This study is limited by the number of workshops run, both in-person and virtually. With three in-person workshops and three virtual workshops, we are able to show conclusive evidence that participants' engagement with science and their science identity increased over the duration of Rising Stargirls' creative arts-based astronomy workshop. However, data from more workshops would allow us to make robust conclusions on how the increase in participants' responses changes across workshops. Rising Stargirls will continue to collect data as it runs future in-person and virtual workshops and we plan to conduct this analysis in more detail as soon as the data are available.

This study is also limited by the time spent with each group of participants. Rising Stargirls' workshops were about two weeks long on average, totalling about 20 hours. Participants would further benefit from creative arts-based astronomy activities sustained over a longer period of time. Developing confidence and enthusiasm for their science ability is not a one-and-done process. In the future, Rising Stargirls plans to run both in-person and virtual workshops over a longer period of time and develop specific programming for returning participants. Specific programming for returning participants would allow for the development of girls' science identities to be tracked over multiple years, as they continue to engage in creative arts-based astronomy activities.

The final limitation that we identify with regards to this work is that we do not have a control group. An example control group would consist of middle-school students who do not participate in a Rising Stargirls workshop.

\subsection{Conclusion}

In conclusion, this work demonstrates how encouraging middle-school girls to bring their own personal creativity, as expressed through the arts, to the study of astronomy promotes engagement with science and development of a science identity. The data from six separate creative arts-based astronomy workshops exhibits the progression of girls' science identities and their engagement with science. These outcomes underscore the value of nurturing creativity and integrating the arts into astronomy education.
 
\section{Availability of source code and requirements}

Code used to analyze data presented in this work is available upon reasonable request to the authors. 

\section{Declarations}

\subsection{List of abbreviations}
\begin{itemize}
    \item COVID-19 - coronavirus disease of 2019

    \item CSV - comma-separated values

    \item NASA - National Aeronautics and Space Administration

    \item NSF - National Science Foundation

    \item PhD - Doctor of Philosophy

    \item STEAM - science, technology, engineering, arts, and math

    \item STEM - science, technology, engineering, and math

    \item URM - underrepresented minorities, defined in this work as American Indian or Alaska Native, Black or African-American, Hispanic or Latinx, Native Hawaiian or other Pacific Islander.

    \item YWCA - Young Women’s Christian Association 

\end{itemize}

\subsection{Ethical Approval (optional)}
This research project has received ethical approval from the University of California, Irvine Institutional Review Board (IRB) under Protocol Number 2550 with University of California, Santa Barbara as a relying institution. The ethical approval granted by the IRB signifies that this research complies with established ethical standards, ensuring the protection of the rights, well-being, and confidentiality of all human participants involved in the study.

\subsection{Consent for publication}

Not applicable.



\subsection{Competing Interests}

AS and JNH declare a non-financial conflict of interest in relation to the current study. AS holds the copyrighted \textit{Rising Stargirls Teaching and Activity Handbook} and JNH holds the copyrighted \textit{Rising Stargirls Teaching Activity Supplement}. These handbooks have been developed for educational purposes and provide resources to support the learning and engagement of participants in Rising Stargirls' programs. The content of the handbooks are referenced in the context of the study, but this conflict of interest does not involve any financial considerations. 

MS, VV, and KW disclose a non-financial conflict of interest in connection with the current study. They have each been actively engaged in organizing and conducting Rising Stargirls workshops. Their involvement includes planning, facilitating, and developing curricula for these workshops. The knowledge and insights gained from their experience with Rising Stargirls may influence the interpretation of findings or discussions related to the workshops in this study. However, this conflict of interest does not involve any financial interests or considerations.

\subsection{Acknowledgements}

This material is based upon work supported by the National Science Foundation, under Award 1753373, by the Heising-Simons Fondation under Award 2023-4634, and by a Clare Boothe Luce Professorship, supported by the Henry Luce Foundation. JNH acknowledges support by the National Science Foundation under Grant No. DGE-1839285 and NSF PHY-1748958, and by the Gordon and Betty Moore Foundation through Grant No. GBMF7392. We give our deepest gratitude to everyone involved in running the six Rising Stargirls workshops discussed in this work. In particular, we thank Loraine Sandoval Ascencio, Christina Dinh, Helena Garcia Escuerdo, Max Fieg, Patricia Fofie, Daniela Gonzalez, and Mei Lin for their instrumental help in running Rising Stargirls' virtual workshops. MS thanks Professor Laura Tucker for her early guidance in conducting science education research. AS thanks Mary Dussault at the Science Education Department at the Harvard-Smithsonian Center for Astrophysics for science education mentoring and help implementing pre- and post-survey assessment strategies into Rising Stargirls program workshops.

\bibliography{paper-refs}

\end{document}